# Status drives how we cite:
# Evidence from thousands of authors

Misha Teplitskiy[1,5,*], Eamon Duede[2,6], Michael Menietti[3,5], Karim R. Lakhani[3,4,5]

**Affiliations:**
[1]University of Michigan
[2]University of Chicago
[3]Harvard University
[4]Institute for Quantitative Social Science at Harvard
[5]Laboratory for Innovation Science at Harvard
[6]Knowledge Lab, University of Chicago

* Correspondence to: Misha Teplitskiy, tepl@umich.edu. School of Information, 4322 North Quad, 105 S State St, Ann Arbor, MI 48109.

# Abstract

Researchers cite works for a variety of reasons, including some having nothing to do with acknowledging influence. The distribution of different citation types in the literature, and which papers attract which types, is poorly understood. We investigate high-influence and low-influence citations and the mechanisms producing them using 17,154 ground-truth citation types provided via survey by 9,380 authors systematically sampled across academic fields. Overall, 54% of citations denote little-to-no influence and these citations are concentrated among low status (lightly cited) papers. In contrast, high-influence citations are concentrated among high status (highly cited) papers through a number of steps that resemble a pipeline. Authors discover highly cited papers earlier in their projects, more often through social contacts, and read them more closely. Papers' status, above and beyond any quality differences, directly helps determine their pipeline: experimentally revealing or hiding citation counts durings the survey shows that low counts cause lowered perceptions of quality. Accounting for citation types thus reveals a "double status effect": in addition to affecting how often a work is cited, status affects how meaningfully it is cited. Consequently, highly cited papers are even more influential than their raw citation counts suggest.

# Introduction

A principle means by which scientists acknowledge their intellectual debts to prior research is through the practice of citation (Merton 1957, 1973). In turn, administrators and researchers often use citations, or metrics derived from them like journal impact factors and h-indices, as a way to quantify intellectual impact (Abbott et al., 2010; Clauset et al., 2017; McKiernan et al., 2019). In addition to their use in evaluation, scholars often interpret citations as indicators of intellectual influence, leading to studies of the intellectual impact of team size (Wu, Wang, and Evans 2019), the long term intellectual impact of scientific works (D. Wang, Song, and Barabási 2013), the winners of major prizes (Garfield & Welljams-Dorof, 1992), how individuals' propensities to impart influence evolves over time (Sinatra et al. 2016), and many other aspects of the research enterprise (Azoulay et al., 2018; Fortunato et al., 2018).

These uses of citations focus on aggregate citation counts, with the assumption, or hope, that the counts broadly measure intellectual influence. However, scholars have long recognized that citations come in different types. While researchers undoubtedly cite some works to acknowledge significant intellectual contributions to their projects, they also cite works for various other reasons, some of which have little to do with acknowledging influence (Bornmann & Daniel, 2008; Cronin, 1984; Liu, 1993; MacRoberts & MacRoberts, 1996; Nicolaisen, 2007). For example, scholars may cite works to provide their readers with context, or to compare their contributions with the prior literature, or to criticize that literature (Catalini et al., 2015). Some works may even be cited because of "coercion" during the publishing process (Wilhite & Fong, 2012). While all citation types may serve useful epistemic functions (except for, perhaps, coercive cites), citations of certain types may have had no discernible intellectual impact on the work in which they are cited, and may have even been discovered after the work was largely complete.

The distribution of citation types in the research literature and how these types are generated is poorly understood. This paper seeks to answer two foundational questions: (1) *what is the distribution of high- and low-influence citations in the literature and are they concentrated among heavily cited, seminal papers or lightly cited, obscure ones?* and (2) w*hat is the role of a paper's status in driving the way it is cited?* Answers to these questions are important for at least two substantive reasons.

First, identifying the subset of truly influential citations helps to better quantify influence. Given that some citations indicate substantial influence while others may denote no influence at all, how any particular paper's influence relates to its overall citation count depends on the type of citations it tends to attract. For example, if high-influence citations are distributed more-or-less randomly throughout the literature, then ascertaining citation type has little effect on quantifying influence: one could simply multiply a work's citation count by the overall fraction of influential citations, and get what is on average the actual measure of its influence. If, on the other hand, low-influence citations tend to accrue to highly cited works, then their lofty counts may reflect their rhetorical efficiency more than, say, their intellectual impact. For example, sometimes citing an established expert in a particular field performs the function of granting legitimacy to the citer's contribution (Cole & Cole, 1972, p. 368), and clearly this function would not be fulfilled as well by referencing an obscure expert (Latour and Woolgar 2013; Latour 1987). However, if the converse were true and low-influence citations go primarily to lightly cited works, then heavily cited papers would be even more impactful in relative terms than their raw citation counts suggest.

Second, focusing on citation types can help elucidate the connection between scientific quality and recognition, and thereby guide the design of institutions to be more meritocratic and efficient. Although extensive research investigates the connection between quality and citation *counts* (e.g. Wang et al., 2013), the connection between quality and actual influence, as reflected by influential citations, is unclear. Citations can be viewed as the outcome of a "citing pipeline," in which authors need to first become aware of a paper, read it early enough and sufficiently carefully to be able to adjust their own work accordingly, and to acknowledge those debts by citing it (Cronin, 1984; P. Wang & Domas White, 1999; P. Wang & Soergel, 1998). In an idealized world without time constraints, this pipeline would favor high-quality work. A researcher might be aware of all relevant papers, read all of them sufficiently closely, become influenced by the high-quality ones, and cite primarily those. Citation counts would then be synonymous with quality, and incentivize authors to produce high quality papers in the future.

However, as Robert Merton famously argued, high-status works and actors, rather than high-quality ones, can benefit at each step of the recognition pipeline, a process referred to in the literature as the Matthew Effect and cumulative advantage (Merton, 1968). For example, receiving prestigious awards has been shown to boost one's citations in biomedicine (Azoulay et al., 2013), and a model of citing with preferential attachment to already highly cited works fits the data well (D. Wang et al., 2013). We hypothesize that in addition to citation amount, status may affect citation type. Famous papers might be easier to discover in time to be meaningfully incorporated, and might be recommended by colleagues more often, suggesting they are worthy of careful attention. These papers may thus have more opportunities to truly influence, and eventually to be cited as such. Lower status papers might still be cited, but may not be read early enough, or closely enough to ascertain whether they are indeed of lower quality. The citations they amass would serve roles other than acknowledging intellectual debts. Favorable status effects may thus create a divergence between quality and citation influence, above and beyond any effects on citation *amount*. If high status yields more high-influence citations, it can exacerbate the status effect on citing amount, causing a "double status effect."

## Data and methods

Studying citation types at scale has proven difficult due to the cost of ground-truth data on influence. For example, the cost of measuring the quality and impact of a subset of research outputs in the UK using non-bibliometric methods has been estimated at €250 million (Else, 2015). Researchers have instead examined other proxies of influence, such as textual analysis (Gerow et al. 2018), prizes (Li et al. 2019), or recruited third-parties to hand code citations for the function they appear to serve in the text (Jurgens et al., 2018; Moravcsik & Murugesan, 1975; Tahamtan & Bornmann, 2019; Valenzuela et al., 2015). Ultimately, however, only the citing authors themselves have ground-truth knowledge as to whether and to what extent they were influenced by their references, but such data exist only on a small scale (Brooks, 1985).

To address this concern, we construct, to our knowledge, the largest existing dataset of ground-truth citation types, labeled by the citing authors themselves (see Methods for details). The data were collected via a highly personalized survey to randomly sampled corresponding authors of papers published in 2015 in the Web of Science database, with each author being asked about two references they made, published primarily in one of 15 fields. 9380 respondents classified the influence of 17154 references. We measured how influential a reference was with the question *"How much did this reference influence the research choices in your paper?"* Answer choices ranged from 1 (very minor influence) to 5 (very major influence). The answer choices and their distribution overall and by field is displayed in Figure 1.

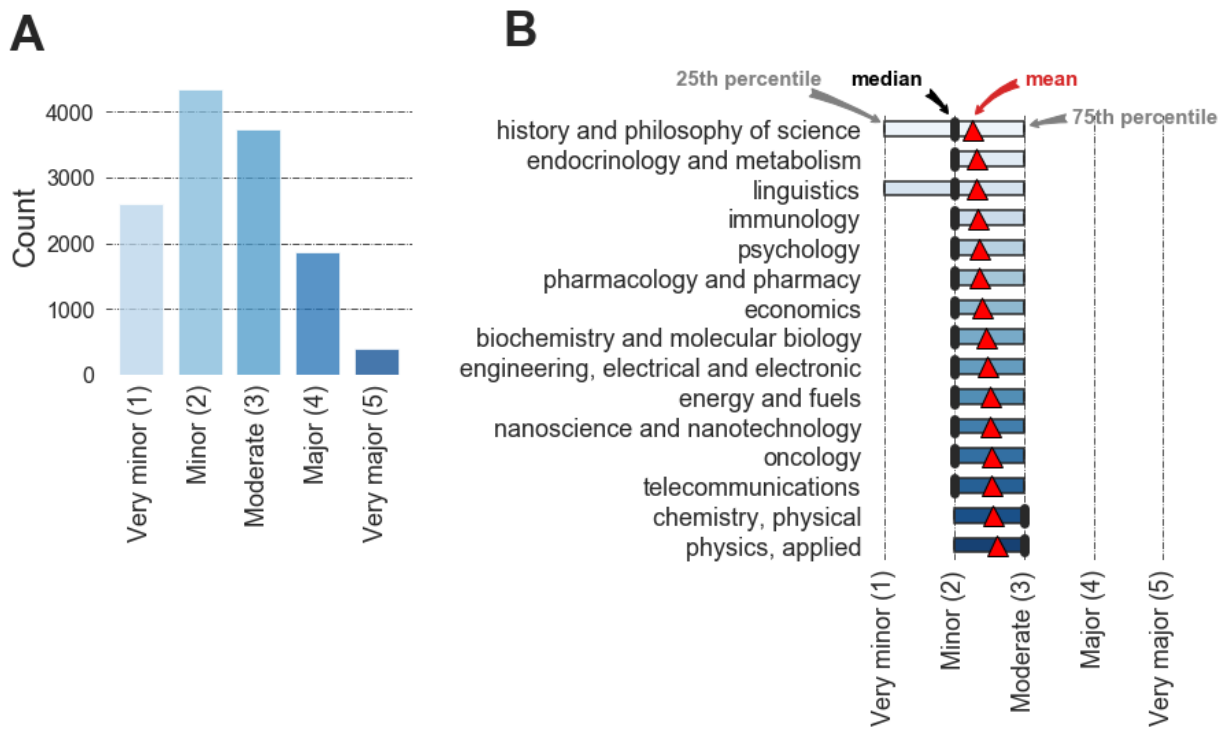

**Fig. 1. A:** Overall distribution of responses, which were described as 1: *Very minor influence (paper would've been very similar without this reference)*, 2: *Minor influence (influenced a small part of paper, e.g. added sentence(s) to Discussion)*, 3: *Moderate influence (influenced an important part of the paper, e.g. additional analysis)*, 4: *Major influence (influenced a core part of paper, e.g. choice of theory or method)*, 5: *Very major influence (motivated the entire project)*. **B:** Boxplot of responses by focal discipline (see Methods). Each box shows the 25th percentile (left edge), the median (heavy black bar), the mean (red triangle), and the 75th percentile (right edge).

Figure 1A shows that the modal citation influence level *Minor (influenced a small part of paper, e.g. added sentence(s) to Discussion (2)*. Overall, 54% of citations were coded as having either *very minor* or *minor* influence. The distribution of citation types varied slightly by field, with only Physical Chemistry and Applied Physics having a higher modal citation type -- *Moderate (influenced an important part of the paper, e.g. additional analysis)*.

To investigate the mechanisms determining citation types, authors were asked to identify when and how they first discovered the reference, and to rate it on several dimensions of quality. Furthermore, surveys were randomly assigned to display (treatment) or hide (control) the reference's citation status, *i.e.* how many times it had been cited and its citation percentile, during normal survey flow (see Appendix: Survey materials). This experiment enables us to establish whether citations can directly change perceptions of quality, which may in turn drive its citation type..

The respondents hailed from all over the world: 19.1% from U.S., 7.3% from China, followed by Italy, India, and Germany, with about 4-5% each. 51.5% of respondents reported employment as full, associate, or assistant professor, and 70.3% identified as male.

# Results

After removing self-citations (7.3%) and cases assigned to the experimental condition (14.9%), we find that the majority of references (53.6%) had at most "minor" influence (1 or 2) on citing authors' research choices. This is consistent with previous literature which argues that the primary goal of research articles is to position and persuade, rather than acknowledge influence (Cozzens, 1989; Gilbert, 1977; Macroberts & Macroberts, 1987; Moravcsik & Murugesan, 1975).

However, citations to works that are highly cited were substantially more influential (per citation) than those to lightly cited papers. Overall, the mean level of influence increases by 0.133 points on a 5-point scale for every 1-unit increase in log-citations (Table S6, model 2 and Fig. 4D). Instead of mean influence, we focus on references that exert *Major (influenced a core part of paper, e.g. choice of theory or method)* or *Very major (motivated the entire project)* influence on their readers, i.e. $\geq 4$. These papers alter the guiding theories or questions of their readers and, sometimes, motivate entirely new projects. Accordingly, we define an indicator variable major_influence and estimate a logistic regression predicting it from the reference's log-citation count, fixed effects for the citing author and field of the reference, and a set of controls consisting of the publication year, whether the corresponding author or a co-author added the reference, and whether it was the first or second reference in the survey (Appendix: Major Influence and Citations, Table S7).

Figure 2 displays the predicted probability that a reference had major influence as a function of its citation count.

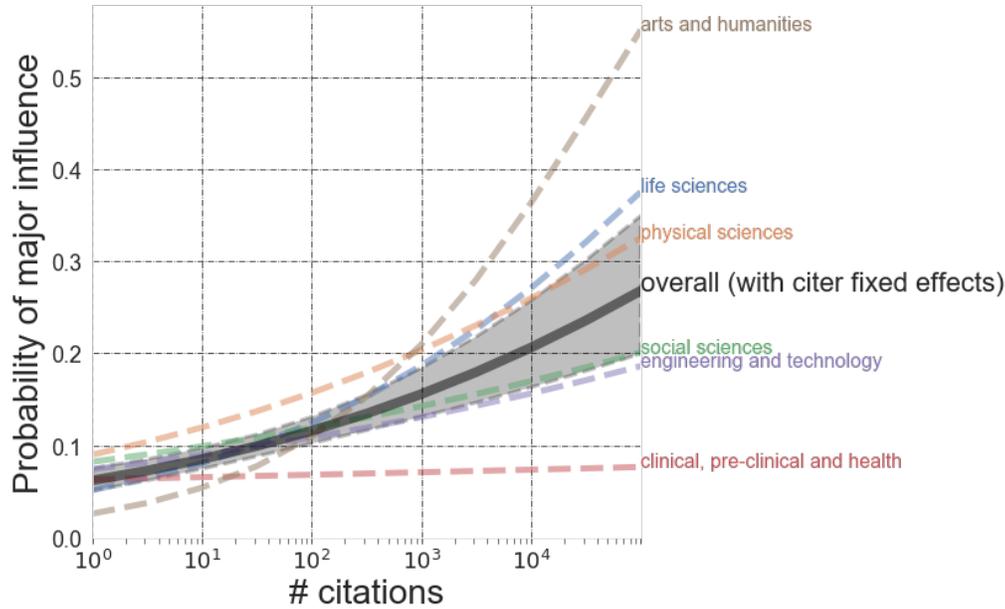

**Fig. 2.** Blue curve with 95% CI shows predicted probability of a reference having major influence (influence >= 4), with fixed effects for the citer and discipline, and controls for year of publication, whether the corresponding author or coauthor added the reference, respondent's expertise in the topics of the reference, and whether it was the first paper in the survey, gray curves show analogous probabilities for each of 6 major research areas.

The curve shows that the probability that a reference had a major influence is low (6.2%) for references with 1 citation but jumps to 15.6% for a reference with 1,000 citations, and 20.6% for a reference with 10,000 citations. That is, citations to highly cited papers are about twice as likely, and to truly famous papers three or more times as likely, to denote major influence. These "within-author" models ensure that the observed differences are not confounded by different citing tendencies or definitions of "influence" across authors.

## Citing pipelines

To better understand the proximate mechanisms that drive highly cited references to have higher per-citation influence, and to answer the more general question concerning the role of status in driving the how we cite, authors were asked when and how they discovered each reference, how well they know it, and for their evaluation of its quality. Drawing on a long-standing distinction in the philosophy of science between "pursuit" and "justification" phases of research (Howard 2006; Richardson 2006; Schickore and Steinle 2006; Hoyningen-Huene 1987; Reichenbach 1938), we hypothesized that researchers' goals change across project lifecycles, giving rise to distinct pursuit and justification citing "pipelines." We expected the earlier research stages of pursuit to focus on the goal of locating the best and most inspirational works for the researcher to invest in reading carefully, and adjust research choices according to the ways in which they are influenced. We expected researchers in the later justification stages to focus on references needed to justify the project and to get it published. These references might be discovered through targeted rather than exploratory search, and not necessitate as careful a reading given that many of the project's key research choices were made earlier.

Empirically, we partitioned project lifecycles into five stages, with the first two constituting pursuit and latter three constituting justification. The pursuit phase consists of *stage 1* - before the project has begun, and *stage 2* - the early period of work (e.g. design, data collection). About 58.2% of cited references were discovered by authors in the pursuit phase. As the project moves into the justification phase, we break the process into *stage 3* - the middle of the project (e.g. analysis), *stage 4* - the finalization of the project (e.g. drafting the manuscript), *stage 5* - the peer review period of the project.

Figure 3 illustrates that reading and citing practices vary systematically and substantially across the pursuit and justification phases. In stages 1 and 2, researchers rely more often on social contacts than in later stages associated with justification, whereas the fraction of references found through database searches, e.g. Google Scholar, increases consistently until peer review (Fig. 2A). Researchers use papers that they deem to be of higher-quality in the pursuit phase than they do in the justification phase, with references dropping in perceived overall quality by more than 10 percentile points from Stages 1 to 4 (Fig 2B). Interestingly, the references discovered during peer review (phase 5) are considered to be of lowest quality overall.

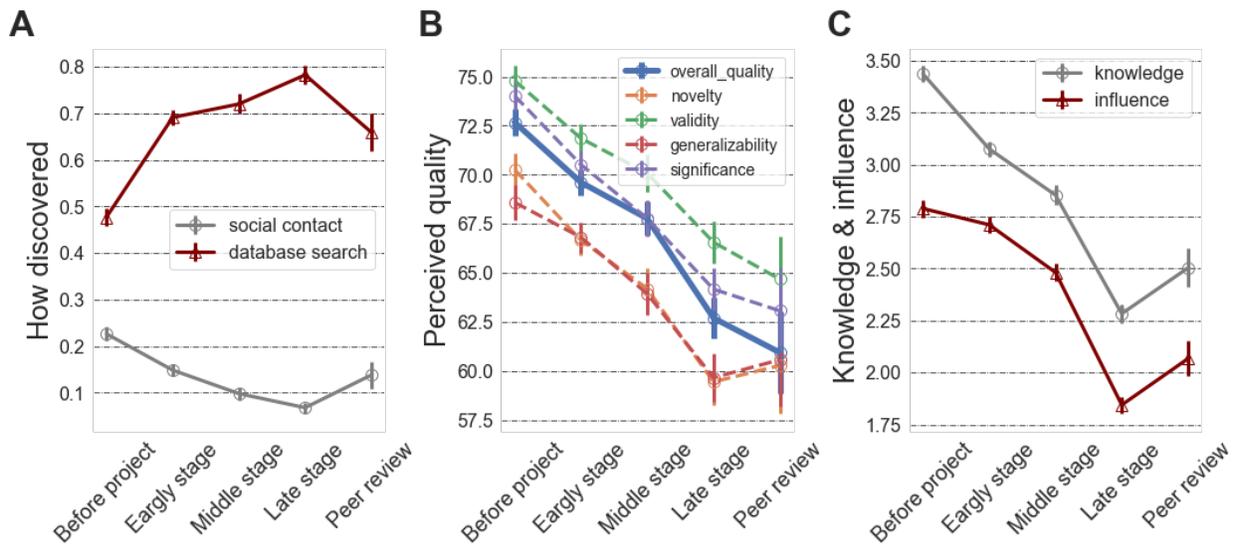

**Fig. 3. A**: How the reference was first discovered by the respondent. "Social contact" combines the response options *"recommended by colleague"* and *"saw in a conference, presentation, or class"*. Responses without at least one of the options checked are excluded from the calculations. **B:** Perceived overall quality (in percentiles) relative to other papers in the field. **C:** Respondent's knowledge of the reference's contents and the influence of the reference on the respondent's research choices (1-5 point scale, 1=lowest).

Fig. 3C shows that authors' knowledge of the contents of the papers they reference, and the influence of those references on the authors, declines across stages. Both decline by about 1 full point on a 5-point scale when a reference from Phase 1 is compared to Phase 4. Not surprisingly, a reference's influence is highly correlated with how well the author knows its content ($\rho=0.573$, $p<0.00$) suggesting that careful reading is a necessary (but not sufficient) condition for influence.

Overall, these patterns reveal a stark difference in discovery, reading, and citing practices across project lifecycles. We use the metaphor of pipelines to reflect the path-dependency that is likely for these practices. Early in the research lifecycle, in the pursuit pipeline, references are more likely to come from low-volume sources that necessitate a high investment, *e.g.* attending presentations. These investments are made carefully, focusing on the highest quality work. Later, in the justification pipeline, more references come from high-volume sources requiring little investment, *e.g.* an internet search. Targeted search suggests that authors have already chosen a research problem and research design, and are increasingly searching for references to fulfill specific, pre-specified functions rather than for inspiration. The justification pipeline of citing is consistent with evidence suggesting that many papers are not read before citing (Simkin & Roychowdhury, 2005) and that about 60% of authors have experienced coercion to add citations in the peer review process (Wilhite & Fong, 2012)

**Status and pipelines**

Next, we show that the papers entering the pursuit pipeline are disproportionately high status, which we measure as log-citations. Figure 4 shows that at each step of the citing process, the more highly cited the paper the more likely it is to be cited through the pursuit pipeline.

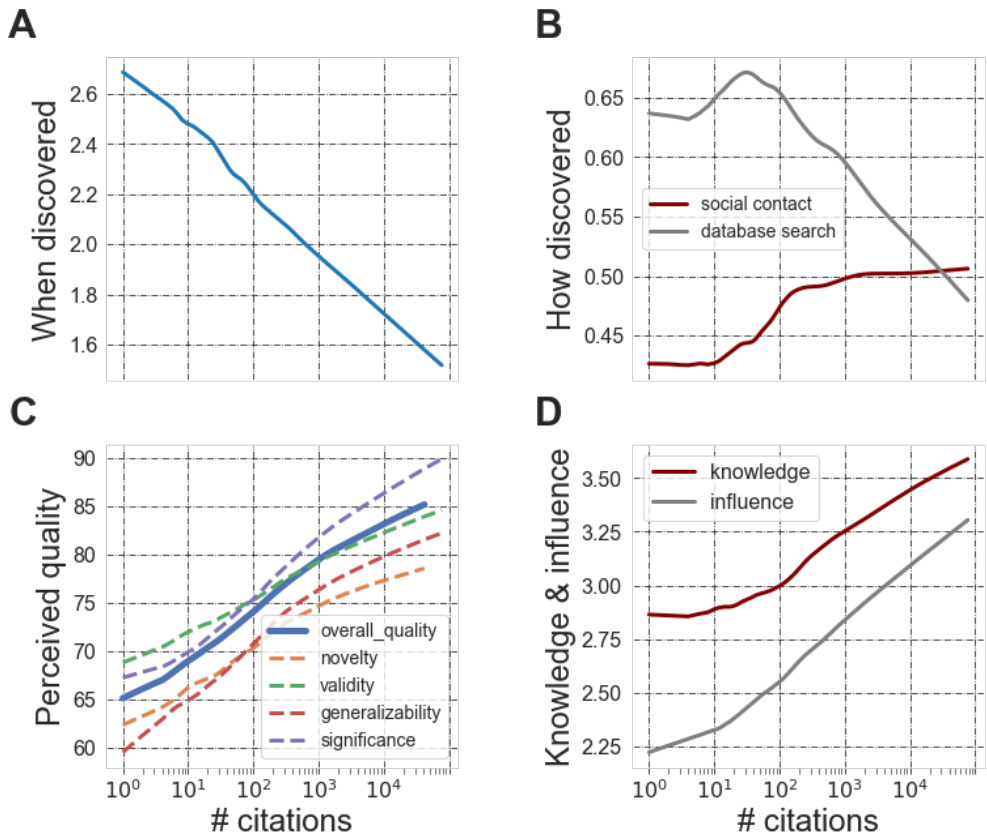

**Fig. 4. A**: When the reference was discovered, using the five phases from Fig. 3 with 1=*Before the project* and so on. **B**: How the reference was first discovered by the respondent. *Social contact* consists of "recommended by colleague" and "saw in presentation." Responses without at least one of the options checked are excluded from the calculations. **C**: Perceived quality (in percentiles)

relative to other papers in the field. **D**. Respondent's knowledge of the reference's contents and the influence of the reference on the respondent's research choices (1-5 point scale, 1=lowest).

Relative to lightly cited papers, highly cited, seminal papers are discovered earlier in the research cycle (Fig 4A), and more often through social contact (Fig 4B). More highly cited papers are perceived to be of higher quality (Fig. 4C), are read more carefully by citing authors and cited as more influential (Fig. 4D).

Overall, Figures 3 and 4 show strong associations that suggest that the high- and low-influence citation types are outcomes of pursuit and justification pipelines, and that the papers in the pursuit pipeline are disproportionately high status.

## Status and quality

A plausible causal chain connecting these pipeline stages turns on the quality of a paper. Quality leads researchers to read a paper earlier and more carefully (and to share it with colleagues), which leads to more influence, which, in turn, leads to a more influential citation. However, a paper's quality may not be readily apparent, and require some reading just to assess. Yet researchers read relatively few papers, and even fewer carefully (Simkin & Roychowdhury, 2005; Tenopir et al., 2015). It is thus crucial to understand where *perceptions* of quality originate.

Earlier literature assumed that citations simply reflect rather than cause quality perceptions (Smith, 1981, pp. 84–85). We hypothesized that researchers would use citation counts as a heuristic for quality, resulting in reverse causation. To test this hypothesis, we experimentally varied whether citation information was shown (treatment) or hidden (control) as respondents took the survey. The treatment information consisted of the reference's true citation count and ranking in the field-year-specific citation distribution, and was displayed before respondents rated the reference on 5 dimensions of quality: overall quality, validity, significance, generalizability, and novelty (see Appendix: Survey materials). We expected showing citations to affect low-cited works negatively and highly cited works positively. The associations between citation percentile and perceived overall quality for control (gray loess curve) and treatment (red loess curve) observations are displayed in Fig. 4A, and support the negative effect. Fig 4B quantifies this pattern by calculating treatment effects on all five dimensions of quality for papers in the bottom-50% and top-50% of their citation distributions (see Appendix: Status signal experiment).

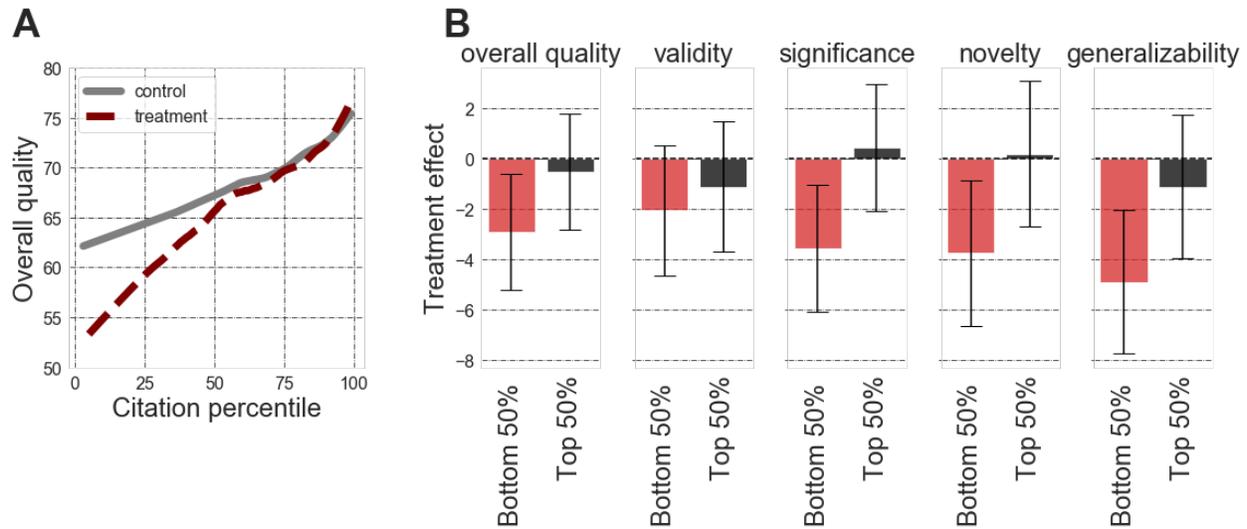

**Fig. 5**. **A**. Scatter plot of perceived "overall quality" and citation percentile of control (gray) and treatment (red) observations, with loess curves. **B**. Treatment effects of status signal on five dimensions of perceived quality. The effects are estimated separately for references below the median in discipline-year citation distribution ("Bottom 50%") and above it ("Top 50%"), using only references published in the 15 focal fields in the 3 focal years (See Data and Methods).

Even this "light-weight" status signal produced a consistent pattern: exposing citation counts caused perceptions of *overall quality, significance, generalizability*, and *novelty* to fall significantly for the bottom half of papers. The effect on perceived *validity* matches others in direction, but does not reach statistical significance, perhaps because citing authors feel more certain of their ability to assess validity. Meanwhile, exposing citations had no substantial effect on papers in the top half. If treatment effects are modeled as linear, then showing citation information harms the perceived "overall quality" for the bottom ~90% of papers (see Appendix: Status signal experiment). Given that the respondents had already cited the references in question, it is plausible that, for many, the citation information was only a reminder. This leads us to speculate that the observed effects are, in fact, larger in a population seeing the references for the first time.

This experiment shows the role of status (citation count and rank) in generating quality perceptions. Papers with many citations are perceived more favorably, above and beyond any underlying quality differences. Favorable perceptions are in turn associated with careful reading and influence. If improved quality perceptions lead to yet more citing, the strength of the heuristic can self-reinforce (DiPrete & Eirich, 2006). It is worth noting, however, that in a world where citations did not serve this heuristic role and simply reflected quality, we would likely still see the same associations observed in Figures 3A-B and 4A-C. However, in such a world, we would expect researchers to read all of their references in some depth before in order to determine their quality. Yet Fig. 3C and 4D shows that is not the case, as researchers do not know many of their references beyond the content of the abstract. Our signaling experiment helps to explain this discrepancy -- researchers to some extent assume quality from citations, and presumably focus reading effort on only the best-citations papers.

# Discussion

In summary, using large-scale ground-truth data on citation types, sampled systematically across all branches of science and humanities, as well as an experiment, we were able to address the two fundamental research questions concerning citation types.

First, we found that less than half of all citations denote meaningful influence. The truly influential works can be systematically identified in authors' reference lists: they are the most highly cited papers. Not only do the most highly cited papers receive a large share of all citations, but the type of citation they tend to accrue are those that are more influential, having led researchers to change the problems they pursue and how they pursue them. A key practical implication of this finding is that citation counts underestimate the relative intellectual impact of the most highly cited papers. For instance, a single paper with 1,000 citations is likely to be much more influential than a portfolio consisting of 100 papers with 10 citations each. This evidence reveals an even starker inequality in influence than has been acknowledged and undermines the theory that the research frontier is advanced primarily by the combined efforts of average researchers rather than by a small number of elites known as the Ortega Hypothesis (Cole & Cole, 1972). While previous work has established that just 10% of most highly cited papers garner as much as 50% of all citations (Bornmann & Leydesdorff, 2017), our findings show that this is a significant underestimate of their share of actual influence.

Second, we find that the processes that drive low- and high-influence citation types resemble two pipelines, with the path a paper is on being set early, and in part by their status. Highly cited, seminal papers disproportionately enter the pursuit pipeline: they are discovered through social connections, are encountered early enough in the lifecycles to change readers' research choices, considered to be of higher quality, and are read more carefully. More obscure and lightly cited papers, if cited at all, disproportionately enter the justification pipeline -- they are discovered more often through database searches, later in project lifecycles, considered to be of lower quality, and read less carefully. Admittedly, these associations between citation status and each pipeline step are correlational. Yet using a randomized controlled trial, we find that one step, the association between citation status and perceptions of quality, is causal: low citation counts cause researchers to view those papers as being of lower quality overall. Although we use this experiment in the context of citation types, it presumably applies more generally, such as to raw citation counts and bears on questions concerning cumulative advantage. Overall, future research is needed to test whether these associations, plausibly causal, are indeed so.

In sum, famous papers are even more important than previously recognized, and receive preferential treatment during each step of the citing process. However, it would be a mistake to conclude that the more fundamental mechanism is simply that prominent papers are of higher underlying quality, and therefore attract outsized attention and influence. In fact, in the absence of clear status signals, authors do not perceive large quality differences in the papers they reference (Fig. 5A, gray curve). On the contrary, our work points to the existence of "double status effects." Status signals in the form of citations and ranks amplify those differences, resulting in perceptions that more closely conform to the huge differences in citation outcomes (Fig. 5A, red curve). This translation of small differences in quality into large differences in outcomes such as influence is consistent with the literature on cumulative advantage in science, often referred to as the Matthew Effect (Allison et al., 1982; Azoulay et al., 2013; Bol et al., 2018; Merton, 1968) and our experiment provides the first truly experimental evidence in support of the Matthew Effect.

A practical implication of these findings concerns the common practice of displaying citation counts alongside the papers in search and discovery contexts. While it may not be surprising that this practice amplifies existing differences in perceived quality, our experiment suggests that "losers" in these systems may greatly outnumber the "winners" 9-to-1. It is important for future research to compare the welfare benefits of what is presumably more efficient search against the costs for decreased visibility of the vast majority of scholarly work.

Lastly, our approach adds a new, powerful method to the toolkit of measuring quality and impact in science, one which has been called for by scholars (Azoulay et al., 2018; Clauset et al., 2017): large-scale surveys and survey experiments. Experiments in particular enable scholars to reach clean, credible conclusions regarding scientific practices, while large scale surveying enables stakeholders primarily interested in accurate measurement to solicit unusually deep and wide information on the impact of their programs, scholars, and institutions.

# References

Abbott, A., Cyranoski, D., Jones, N., Maher, B., Schiermeier, Q., & Van Noorden, R. (2010). Metrics: Do metrics matter? *Nature*, *465*(7300), 860–862. https://doi.org/10.1038/465860a

Allison, P. D., Long, J. S., & Krauze, T. K. (1982). Cumulative Advantage and Inequality in Science. *American Sociological Review*, *47*(5), 615–625. JSTOR. https://doi.org/10.2307/2095162

Azoulay, P., Graff-Zivin, J., Uzzi, B., Wang, D., Williams, H., Evans, J. A., Jin, G. Z., Lu, S. F., Jones, B. F., Börner, K., Lakhani, K. R., Boudreau, K. J., & Guinan, E. C. (2018). Toward a more scientific science. *Science*, *361*(6408), 1194–1197. https://doi.org/10.1126/science.aav2484

Azoulay, P., Stuart, T., & Wang, Y. (2013). Matthew: Effect or Fable? *Management Science*, *60*(1), 92–109. https://doi.org/10.1287/mnsc.2013.1755

Barabási, A.-L., Song, C., & Wang, D. (2012). Handful of papers dominates citation. *Nature*, *491*(7422), 40–40. https://doi.org/10.1038/491040a

Bol, T., Vaan, M. de, & Rijt, A. van de. (2018). The Matthew effect in science funding. *Proceedings of the National Academy of Sciences*, *115*(19), 4887–4890. https://doi.org/10.1073/pnas.1719557115

Bornmann, L., Anegón, F. de M., & Leydesdorff, L. (2010). Do Scientific Advancements Lean on the Shoulders of Giants? A Bibliometric Investigation of the Ortega Hypothesis. *PLOS ONE*, *5*(10), e13327. https://doi.org/10.1371/journal.pone.0013327

Bornmann, L., & Daniel, H. (2008). What do citation counts measure? A review of studies on citing behavior. *Journal of Documentation*, *64*(1), 45–80. https://doi.org/10.1108/00220410810844150

Bornmann, L., & Leydesdorff, L. (2017). Skewness of citation impact data and covariates of citation distributions: A large-scale empirical analysis based on Web of Science data. *Journal of Informetrics*, *11*(1), 164–175. https://doi.org/10.1016/j.joi.2016.12.001

Brooks, T. A. (1985). Private acts and public objects: An investigation of citer motivations. *Journal of the American Society for Information Science*, *36*(4), 223–229. https://doi.org/10.1002/asi.4630360402

Catalini, C., Lacetera, N., & Oettl, A. (2015). The incidence and role of negative citations in science. *Proceedings of the National Academy of Sciences*, 201502280. https://doi.org/10.1073/pnas.1502280112

Clauset, A., Larremore, D. B., & Sinatra, R. (2017). Data-driven predictions in the science of science. *Science*, *355*(6324), 477–480. https://doi.org/10.1126/science.aal4217

Cole, J. R., & Cole, S. (1972). The Ortega Hypothesis. *Science*, *178*(4059), 368–375. https://doi.org/10.1126/science.178.4059.368

Cozzens, S. (1989). What do citations count? The rhetoric-first model. *Scientometrics*, *15*(5–6), 437–447. https://doi.org/10.1007/BF02017064

Cronin, B. (1984). The Citation Process. *The Role and Significance of Citations in Scientific Communication*, *103*.

DiPrete, T. A., & Eirich, G. M. (2006). Cumulative Advantage as a Mechanism for Inequality: A Review of Theoretical and Empirical Developments. *Annual Review of Sociology*, *32*(1), 271–297. https://doi.org/10.1146/annurev.soc.32.061604.123127

Else, H. (2015, July 13). *REF 2014 cost almost £250 million*. Times Higher Education (THE).

https://www.timeshighereducation.com/news/ref-2014-cost-250-million

Fortunato, S., Bergstrom, C. T., Börner, K., Evans, J. A., Helbing, D., Milojević, S., Petersen, A. M., Radicchi, F., Sinatra, R., Uzzi, B., Vespignani, A., Waltman, L., Wang, D., & Barabási, A.-L. (2018). Science of science. *Science*, *359*(6379). https://doi.org/10.1126/science.aao0185

Garfield, E., & Welljams-Dorof, A. (1992). Of Nobel class: A citation perspective on high impact research authors. *Theoretical Medicine*, *13*(2), 117–135. https://doi.org/10.1007/BF02163625

Gilbert, G. N. (1977). Referencing as Persuasion. *Social Studies of Science*, *7*(1), 113–122.

Jurgens, D., Kumar, S., Hoover, R., McFarland, D., & Jurafsky, D. (2018). Measuring the Evolution of a Scientific Field through Citation Frames. *Transactions of the Association for Computational Linguistics*, *6*, 391–406. https://doi.org/10.1162/tacl_a_00028

Liu, M. (1993). Progress in documentation the complexities of citation practice: A review of citation studies. *Journal of Documentation*, *49*(4), 370–408. https://doi.org/10.1108/eb026920

Macroberts, M. H., & Macroberts, B. R. (1987). Testing the Ortega hypothesis: Facts and artifacts. *Scientometrics*, *12*(5), 293–295. https://doi.org/10.1007/BF02016665

MacRoberts, M. H., & MacRoberts, B. R. (1996). Problems of citation analysis. *Scientometrics*, *36*(3), 435–444. https://doi.org/10.1007/BF02129604

McKiernan, E. C., Schimanski, L. A., Muñoz Nieves, C., Matthias, L., Niles, M. T., & Alperin, J. P. (2019). Use of the Journal Impact Factor in academic review, promotion, and tenure evaluations. *ELife*, *8*, e47338. https://doi.org/10.7554/eLife.47338

Merton, R. K. (1968). The Matthew Effect in Science: The Reward and communication system of science. *Science*, *199*, 55–63.

Moravcsik, M. J., & Murugesan, P. (1975). Some Results on the Function and Quality of Citations. *Social Studies of Science*, *5*(1), 86–92. https://doi.org/10.1177/030631277500500106

Nicolaisen, J. (2007). Citation analysis. *Annual Review of Information Science and Technology*, *41*(1), 609–641. https://doi.org/10.1002/aris.2007.1440410120

Radicchi, F., Weissman, A., & Bollen, J. (2017). Quantifying perceived impact of scientific publications. *Journal of Informetrics*, *11*(3), 704–712. https://doi.org/10.1016/j.joi.2017.05.010

Simkin, M. V., & Roychowdhury, V. P. (2005). Stochastic modeling of citation slips. *Scientometrics*, *62*(3), 367–384. https://doi.org/10.1007/s11192-005-0028-2

Smith, L. C. (1981). Citation Analysis. *Library Trends*, *30*(1), 83–106.

Tahamtan, I., & Bornmann, L. (2019). What do citation counts measure? An updated review of studies on citations in scientific documents published between 2006 and 2018. *Scientometrics*, *121*(3), 1635–1684. https://doi.org/10.1007/s11192-019-03243-4

Tenopir, C., King, D. W., Christian, L., & Volentine, R. (2015). Scholarly article seeking, reading, and use: A continuing evolution from print to electronic in the sciences and social sciences. *Learned Publishing*, *28*(2), 93–105. https://doi.org/10.1087/20150203

Valenzuela, M., Ha, V., & Etzioni, O. (2015). Identifying Meaningful Citations. *AAAI Workshop: Scholarly Big Data*.

Van Noorden, R. (2014). Scientists may be reaching a peak in reading habits. *Nature News*. https://doi.org/10.1038/nature.2014.14658

Wang, D., Song, C., & Barabási, A.-L. (2013). Quantifying Long-Term Scientific Impact.

  *Science*, *342*(6154), 127–132. https://doi.org/10.1126/science.1237825

Wang, P., & Domas White, M. (1999). A cognitive model of document use during a research project. Study II. Decisions at the reading and citing stages. *Journal of the American Society for Information Science*, *50*(2), 98–114. https://doi.org/10.1002/(SICI)1097-4571(1999)50:2<98::AID-ASI2>3.0.CO;2-L

Wang, P., & Soergel, D. (1998). A cognitive model of document use during a research project. Study I. Document selection. *Journal of the American Society for Information Science*, *49*(2), 115–133. https://doi.org/10.1002/(SICI)1097-4571(199802)49:2<115::AID-ASI3>3.0.CO;2-T

Wilhite, A. W., & Fong, E. A. (2012). Coercive Citation in Academic Publishing. *Science*, *335*(6068), 542–543. https://doi.org/10.1126/science.1212540

**Acknowledgments:** We thank seminar participants at the University of Michigan School of Information, MIT Sloan School of Management, NESTA Innovation Growth Lab, Harvard Kennedy School Growth Lab, University of Chicago Knowledge Lab, International Conference on Computational Social Science, International Conference on Science and Technology Indicators, International Conference on Scientometrics & Informetrics.

**Funding:** MacArthur Foundation Research Network on Opening Governance; Schmidt Futures #785, University of Chicago's BIG Ideas Generator

**Author contributions:** MT was involved in conceptualization, methodology, data analysis, and writing. ED was involved in conceptualization, data curation, and writing. MM was involved in methodology, data analysis, and writing. KRL was involved in conceptualization, methodology, and writing.

**Competing interests:** Authors declare no competing interests

**Data and materials availability:** The nature of the data preclude full anonymization. For example, a citation count > 10,000 can identify specific references (and potentially their citers) with relative ease. However, an anonymized version of a subset of the dataset that is sufficient to reproduce the key findings can be made available upon request.

# Methods

We used rich data from the complete Clarivate Analytics *Web of Science* (WoS) data through 2015 to systematically sample the literature and survey the scientific community. The *Web of Science* attributes journals to disciplines, and disciplines to six major subjects - *Arts and Humanities*, *Clinical, Pre-clinical & Health*, *Engineering & Technology*, *Life Sciences*, *Physical Sciences*, and *Social Sciences*. We could not survey all authors in all disciplines, and instead used the following procedure to choose a subset of disciplines. In each subject we sought to identify disciplines with high coverage by WoS and where citation-based metrics were likely to be salient. Accordingly, for each discipline, we averaged impact factors (IF) of its top five journals, and then ranked disciplines according to this average. For each of the six major subjects, we took the 2-3 highest average-IF disciplines. The selected disciplines were *biochemistry & molecular biology, physical chemistry, economics, endocrinology & metabolism, energy & fuels, electrical & electronic engineering, history & philosophy of science, immunology, linguistics, nanoscience & nanotechnology, oncology, pharmacology & pharmacy, applied physics, psychology, and telecommunications*.

For each discipline, we identified all research articles published in the years 2000, 2005, and 2010 and ranked them according to the number of citations they had accrued through the year 2015. Uncited items were included in the citation distribution. For each percentile of this discipline-year specific distribution, we randomly selected five papers (cited papers), and five random papers citing each of those papers in 2015 (citing papers). If five citing papers were not available, we randomly selected other papers in that percentile and repeated the procedure. The corresponding author of each citing paper for whom WOS had an email address was contacted with a personalized survey (see *Supplementary Materials: Materials and Methods* for details) and asked about two references. Two references were used in order to enable author-fixed effects statistical modeling. The first (focal) reference was selected as above, and the second reference was chosen from the same citing paper, if possible in the same discipline and publication year as the first. If such a second reference was not available, the restrictions were loosened until a suitable second reference could be found, or it was chosen at random. The selection procedure resulted in the set of second papers being on average more highly cited than the set of first papers.

We sent email solicitations to 63,049 corresponding authors of the citing papers. From this risk set, 20.2% (n=12670) of the recipients opened the provided link, and about 15.0% (n=9425) reached the last page. About 10% of emails were undeliverable. Male authors were more likely to reply, while authors publishing in high impact journals were slightly less likely to reply (Appendix: Non-response analysis). In analyses, we removed self-citations (~8%) from the dataset because authors have been shown to evaluate their own papers much more positively than others' papers (Radicchi et al., 2017). Self-citations were self-reported via a checkbox. Although self-reports are imperfect, self-citation bibliometric databases like WOS have substantial identity disambiguation problems, which can lead to many false positive self-citations. For robustness, we repeated our analysis with WOS-based self-citations and found qualitatively similar results (available upon request).

Respondents were asked to evaluate the references on several dimensions of quality (overall quality, validity, novelty, generalizability, and significance) with the question, "Rate this reference against others in the field on the following characteristics, with 50th percentile denoting the typical paper" (Fig. S4).

We experimentally manipulated the information respondents observed before evaluating papers. The control (85%) and treatment (15%) forms were identical except the treatment form displayed the following status signal, "Our records indicate that this paper has been cited X time(s), which ranks it in the [top/bottom] Y percentile among all papers published in the field in [year of publication]," where X was the paper's true citation count in Web of Science in 2015 and Y was the true percentile in the citation distribution (Appendix: Survey materials). If citation counts affect only the number of people who see or read a paper but do not change their perceptions of it, then exposing citation counts should not affect respondents' ratings. On the other hand, if the treatment changes perceptions, it may have opposite effects when citation counts are relatively low vs. high. High citation counts, indicated by the count and the word "top" in "top Y percentile", constitute a positive signal that may improve perceptions and vice versa for counts in the "bottom" percentiles.

# Supplementary Information for

# "Status drives how we cite: Evidence from thousands of authors"

**Contents**

# Survey materials

The following text and figures illustrate the survey flow. Figure S1 displays the (anonymized) recruitment email.

[ Figure S1 about here ]

After clicking on the link, respondents proceeded to confirm that the paper was indeed theirs and read IRB information. Next, they proceeded to a randomized page, the two versions of which are displayed in Figure S2. The control (panel A) and treatment (panel B) versions are identical except that treatment includes the reference's citation information.

[ Figure S2 about here ]

Next, respondents answered questions about their knowledge of the reference, how much it influenced them, which aspects of their work were influenced (Figure S3). To account for ordering effects in answer choices, respondents were randomized into two forms with identical questions but reversed answer choice order. Form A's answer choices ranged from smallest/least to biggest/most, while form B had the opposite ordering. Next, respondents rated the reference on various dimensions of quality (Figure S4). Lastly, respondents provided some demographic information.

[ Figure S3 about here ]

[ Figure S4 about here ]

# Nonresponse analysis

*Disciplines*

Response rates were measured by clicks on the personalized survey link. Rates varied substantially across disciplines. The lowest response rate came from oncology (12.9%) and the highest (34.1%) from history and philosophy of science. The number of completed responses and response rates by discipline are displayed in Figure S5.

[ Figure S5 about here ]

*Impact factor, gender, and status signal*

A response was defined as clicking on the personalized Qualtrics link (*opened.survey*=True). Completing the survey was defined as exiting out of the last page (*completed.survey*=True). To investigate response rates by impact factor of the respondent's journal, gender, and assignment to status signal condition, we estimate a logistic regression with *opened.survey* as the outcome regressed on the *Web of Science* 2015 impact factor, a 3-level gender variable {"Unknown", "Male", "Female"}, and status signal indicator. The model was specified as follows:

$$\log {Y_{ij}}/{1 - Y_{ij}} = \beta_0 + \beta_1 author.female + \beta_2 author.male + \beta_3 source.impact.factor + \beta_4 status.signal + \epsilon_{ij}$$
(1)

Where $Y_{ij}$ is either *opened.survey* or *completed.survey*. Estimates from this regression are displayed in Table S1, model 1 and average marginal effects are displayed in Table S2, model 1. Author gender was inferred using Genderize.io (https://genderize.io/). Average marginal effects indicate that authors of papers in high impact journals were somewhat less likely to respond, with response rate decreasing by 0.73% per unit impact factor, while "Female" and "Male" authors were 2.2% and 5.7%, respectively, more likely to respond than "Unknown." As expected, the randomly assigned status signal did not significantly affect response rates.

[ Table S1 about here ]

[ Table S2 about here ]

Of more concern is whether the status signal affected survey completion rates. Estimates from a logistic regression predicting *completed.survey*=True are displayed in Table S1: Model 2. The point estimate for the average marginal effect of *status.signal* on completion is 0.0071 (0.71%, $SE = 0.004$, $p = 0.071$). Thus we conclude that exposing citation signals increased completion rates by a very small, if any, amount.

## Status signal experiment

*Average treatment effect*

Status can be high or low, and we expect high status signals to improve perceptions of quality, and low signals to harm them. The expected heterogeneity of the treatment makes the average treatment effect (ATE) a poor summary of it. For completeness, we provide the ATE in Table S3 below, estimated with the following regression specifications.

$$attribute_{ij} = \beta_0 + \beta_1 status.signal_i + \beta_2 X_{ij} + \epsilon_{ij}$$
(2)

In this specification, *attribute$_{ij}$* is the rating of a quality attribute by author *i* of reference *j*, *status.signal$_i$* is an indicator of whether author *i* received a status signal (assignment to treatment was done at the respondent, not reference, level) and $\beta_1$ measures the ATE, and $X_{ij}$ is a set of control variables describing author *i* and reference *j*. Control variables are indicators for the author's gender and academic position, and indicators for the reference's discipline, publication year, whether it was added by the responding author or a co-author, and whether it appeared first or second in the survey.

[ Table S3 about here ]

Most columns (attributes) in Table S3 show the ATE of *status.signal* to be negative, with effects on perceived validity and generalizability reaching significance.

*Heterogeneous treatment effects*

We now present two ways to examine treatment effect heterogeneity. First, we partition the field-year-specific citation distribution into halves, above and below the median. The break at the median is natural because at that point the presentation of the status signal changes qualitatively, from "bottom *X*% of the citation distribution" to "top X% of the citation distribution." For this analysis we exclude observations right at the median, where the status signal presentation didn't include the words "bottom" or "top." The specification is

$$attribute_{ij} = \beta_0 + \beta_1 above.median_{ij} + \beta_2 status.signal_i + \beta_3 above.median_{ij} * status.signal_i + \beta_4 X_{ij} + \epsilon_{ij}$$

(3)

where *attribute*$_{ij}$ is the rating of a quality attribute by citer *i* of paper *j*, $\beta_0$ is the mean attribute for bottom-50% papers in control condition, *above.median*$_{ij}$ an indicator variable for top-50%, *status.signal*$_i$ is an indicator for treatment, $X_{ij}$ is a set of controls of the citer and paper, and $\epsilon_{ij}$ is the error. Errors are clustered by respondents, and the controls consist of the paper's discipline and publication year, the citer's gender and position, and whether the respondent or co-author added the reference.

For papers below median citations, showing the status signal tends to harm quality perceptions, as indicated by negative and statistically significant coefficients of *status.signal*. The effect on perceptions of quality is less precisely estimated, but is consistent with the other attributes. For papers above the median, the effect of the status signal is positive for all dimensions, although not consistently statistically significant.

[ Table S4 about here ]

Finally, we model attribute ratings as linear functions of citation percentile (field-year specific), with a separate intercept and slope for control and treatments observations. The specification takes the same form as (3) with *above.median*$_{ij}$ replaced by *percentile*$_{ij}$. Estimates from this regression are displayed in Tables S5.

[ Table S5 about here ]

The estimates show a consistent pattern: the status signal exaggerates the underlying (control) relationship. The status signal harms the perception of low-cited papers (coefficients of *status.signal* are negative and, with the exception of "validity," statistically significant. However the penalty gets smaller as citation counts, and therefore the nature of the signal, improve - coefficients of the interaction effect are positive and, with the exception of "validity," statistically significant.

To locate the approximate point in the citation distribution at which the status signal becomes a net positive for perceptions, we locate the percentile at which the control and status-signal lines intersect. For "overall quality" that percentile is $p^* = 4.579/0.054 = 89.8 \approx 90$. Thus, for all but approximately top 10% highest cited references, the status signal had a negative effect on perceived quality.

## Influence, knowledge, and citations

We use linear mixed models to quantify the relationship between a reference's citation count and its influence on the corresponding author's knowledge of it and its influence on research choices. Author-fixed effects enable us control for all (stable) differences between authors, including their fields. This approach accounts for the possibility that the composition of authors varies significantly across the citation distribution of references. For example, authors citing lowly and highly cited works may have different standards for "influence." We use the following specification for *influence* and similarly for *knowledge*:

$$influence_{ij} = \alpha_i + \beta_0 log.cites + \beta_1 X_{ij} + \epsilon_{ij} \qquad (4)$$

The indices *i* and *j*, enumerate authors and references, respectively. The author fixed effects $\alpha_i$ denote author-specific intercepts. The set of controls $X_{ij}$ is the same as used in (2) above. *log.cites* is base-10 log of the citation count. Estimates from regressions of this form are shown in Table S6. Models 1 and 2 denote models of *influence* with and without author-fixed effects, respectively. Models 3 and 4 denote models of *knowledge* with and without author-fixed effects, respectively.

[ Table S6 about here ]

All models show a robust association between the dependent variable and log-citations. Model (2) shows that, for a given author, the influence on her of a reference is 0.133 points on a 5-point scale higher per unit increase in log-citations. Model (4) shows that, for a given author, his or her knowledge of the reference is 0.127 points on a 5-point scale higher per unit increase in log-citations.

## Major influence and citations

The indicator *major.influence* is defined as 1 if a reference has "major" or "very major" influence, and 0 otherwise. Overall, 17.6% of references met or exceeded the major influence bar. The relationship of a reference's *major.influence* and its *log-citations* was modeled as a generalized linear mixed model with a logit link function. The fixed effects were modeled as (citing) authors within (cited) disciplines of the references. Such a specification ensures that the estimated coefficients are not affected by the highly unequal number of observations per (cited) discipline, nor by the possibly different types of authors who use highly vs lowly cited references. Estimates were done using the *lme4* package in R. Marginal effects were calculated using the *ggeffects* package, using the "simulation" type of standard error, which takes into account uncertainty of the fixed and random effects. Table S7 displays estimates from this model.

[ Table S7 about here ]

The model shows a highly significant positive association between *major.influence* and *log.cites*.

## Descriptive statistics

Tables S8 and S9 display the descriptive statistics and correlation table, respectively, for quantitative variables used in the analysis.

[ Table S8 about here ]

[ Table S9 about here ]

**Fig. S1.**

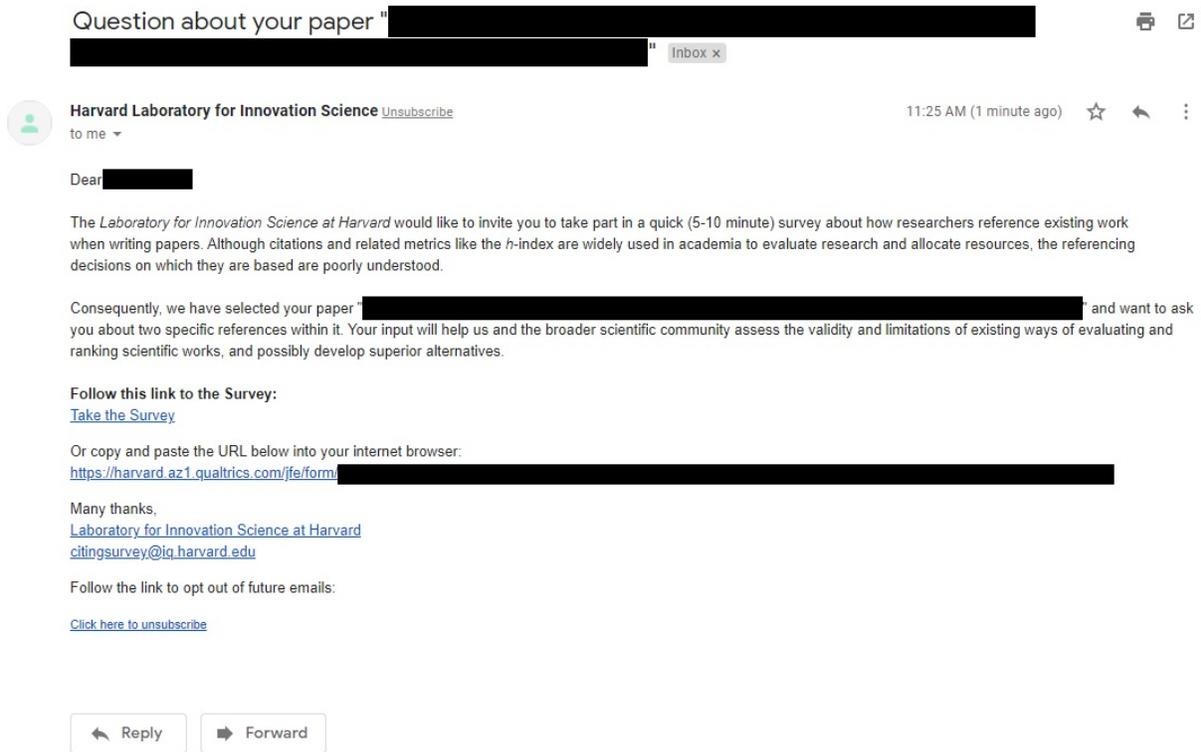

Sample recruitment email.

**Fig. S2.**

A   B

Two forms used for the status signal experiment. 85% of randomly assigned respondents saw the control form (Panel A), which does not show any citation information, and 15% saw the treatment form (Panel B), which displays the true citation count and percentile.

**Fig. S3.**

Reference:

[reference string]

How well do you know this paper?

○ Extremely well (know it as well as my own work)
○ Very well (familiar with all findings, data & methods, all limitations and critiques)
○ Well (familiar with all findings, data & methods, some limitations)
○ Slightly well (familiar with all findings, data & methods)
○ Not well (only familiar with main findings)

How much did this reference influence the research choices in your paper?

○ Very major influence (motivated the entire project)
○ Major influence (influenced a core part of paper, e.g. choice of theory or method)
○ Moderate influence (influenced an important part of the paper, e.g. additional analysis)
○ Minor influence (influenced a small part of paper, e.g. added sentence(s) to Discussion)
○ Very minor influence (paper would've been very similar without this reference)
○ Not sure

Which aspects of your paper did this reference influence? (Mark all that apply)

☐ Only minor influence
☐ Topic
☐ Theory or conceptualization
☐ Data
☐ Methods
☐ Other. Please explain [            ]

Screenshot illustrating questions about the author's knowledge of the reference and its impact on the author. Randomly assigned half of the respondents saw this ordering of answer choices, while another half saw the reverse ordering.

**Fig. S4.**

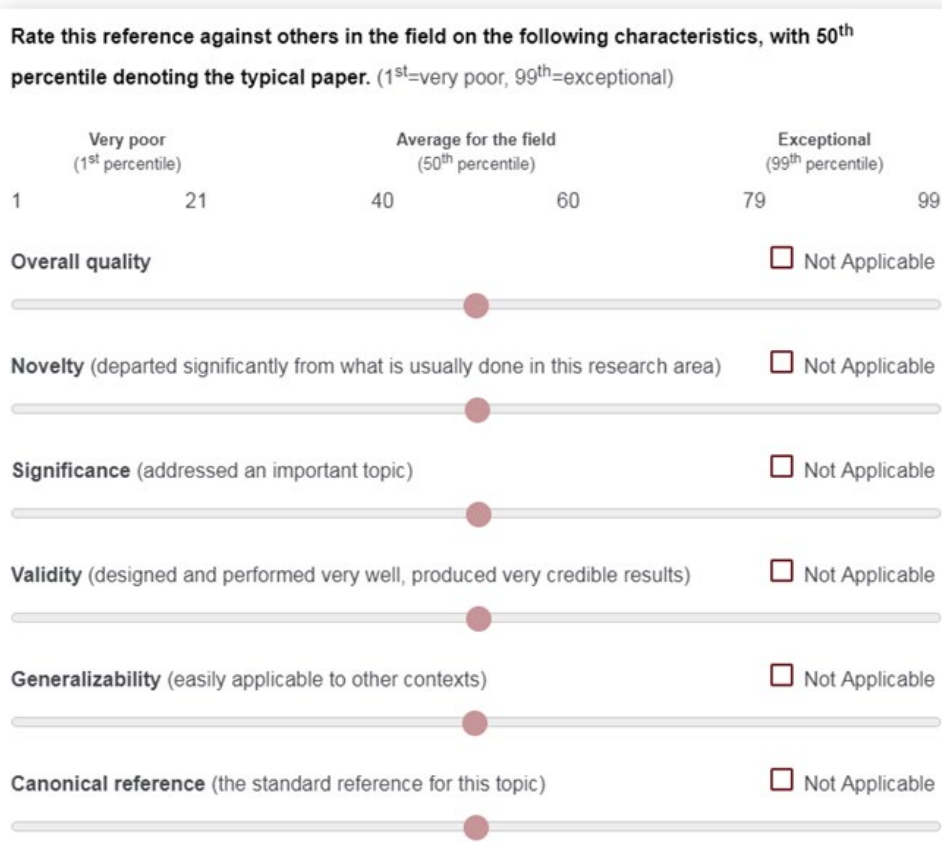

Panel of questions about perceived quality of the reference. The attribute in the last position was randomized to be "Canonical" or "Prominent." Data from this last position is not included in the present analyses due to its only indirect relationship with quality, but is available from the authors upon request.

**Fig. S5**

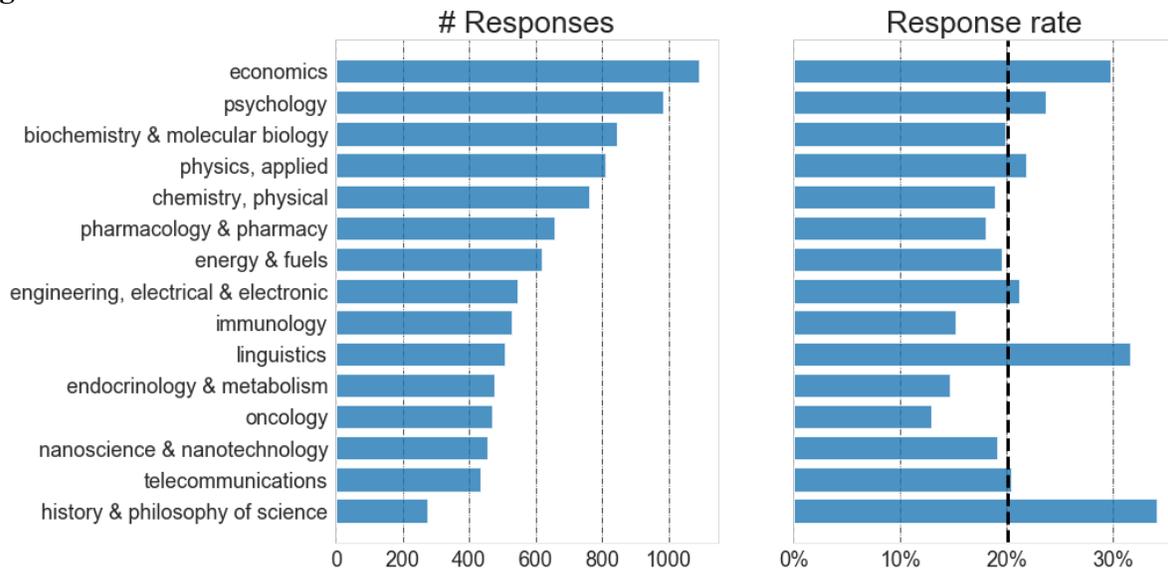

Response counts and response rates by discipline. Each response, if filled out completely, provides data on two references. The dotted line shows the mean response rate.

**Table S1.**

|  | (1) opened_survey | (2) completed_survey |
|---|---|---|
| Intercept | -1.444*** | -1.868*** |
|  | (0.0228) | (0.0260) |
| author.gender = female | 0.134*** | 0.135*** |
|  | (0.0270) | (0.0328) |
| author.gender = male | 0.355*** | 0.427*** |
|  | (0.0240) | (0.0271) |
| source.impact.factor | -0.0458*** | -0.0388*** |
|  | (0.0050) | (0.0055) |
| status.signal | -0.0119 | 0.0560 |
|  | (0.0280) | (0.0310) |
| Observations | 62550 | 62550 |
| df model | 4 | 4 |
| pseudo-R | 0.00518 | 0.00629 |
| LLR p-value | 2.698e-69 | 9.528e-71 |

Estimates from logistic regressions predicting *opening* (Model 1) and *completing* (Model 2) the survey. The reference category for *author.gender* is "Unknown." *p<0.1; **p<0.05; ***p<0.01 for 2-sided Wald tests. Average marginal effects are displayed in Table S2.

**Table S2.**

|  | (1) opened_survey | (2) completed_survey |
|---|---|---|
| author.gender = female | 0.0215*** | 0.0171*** |
|  | (0.004) | (0.004) |
| author.gender = male | 0.0569*** | 0.0541*** |
|  | (0.004) | (0.003) |
| source.impact.factor | -0.0073*** | -0.0049*** |
|  | (0.001) | (0.001) |
| status.signal | -0.0019 | 0.0071 |
|  | (0.004) | (0.004) |

Average marginal effects from logistic regression tables of Table S1. The reference category for *author.gender* is "Unknown." *p<0.1; **p<0.05; ***p<0.01 for 2-sided Wald tests.

**Table S3.**

|  | *Dependent variable:* | | | | |
|---|---|---|---|---|---|
|  | overall.quality | novelty | validity | generalizability | significance |
|  | (1) | (2) | (3) | (4) | (5) |
| status.signal | -1.192* | -0.936 | -1.379* | -2.209** | -0.633 |
|  | (0.607) | (0.741) | (0.678) | (0.756) | (0.631) |
| Controls | Y | Y | Y | Y | Y |
| Observations | 7,376 | 6,848 | 6,866 | 6,715 | 7,359 |
| $R^2$ | 0.063 | 0.035 | 0.036 | 0.036 | 0.057 |
| Adjusted $R^2$ | 0.060 | 0.031 | 0.032 | 0.032 | 0.053 |
| Residual Std. Error | 16.972 (df = 7345) | 19.502 (df = 6817) | 17.799 (df = 6866) | 19.602 (df = 6684) | 17.768 (df = 7328) |

Estimates from OLS regressions of quality attribute ratings on status signal. Robust standard errors clustered at respondent level. + $p<0.1$; * $p<0.05$; ** $p<0.01$; *** $p<0.001$ for two-sided *t*-tests. Constant and controls not shown.

**Table S4.**

|  | *Dependent variable:* | | | | |
|---|---|---|---|---|---|
|  | overall_quality (1) | novelty (2) | validity (3) | generalizability (4) | significance (5) |
|---|---|---|---|---|---|
| above.median | 4.674*** (0.543) | 3.966*** (0.636) | 3.146*** (0.582) | 4.064*** (0.651) | 4.500*** (0.571) |
| status.signal | -2.899* (1.179) | -3.748* (1.479) | -2.058 (1.326) | -4.901*** (1.456) | -3.574** (1.295) |
| above.median X status.signal | 2.379+ (1.346) | 3.942* (1.647) | 0.949 (1.525) | 3.789* (1.659) | 4.012** (1.450) |
| Controls | Y | Y | Y | Y | Y |
| Observations | 7,346 | 6,822 | 6,882 | 6,688 | 7,330 |
| $R^2$ | 0.078 | 0.045 | 0.021 | 0.046 | 0.071 |
| Adjusted $R^2$ | 0.074 | 0.040 | 0.017 | 0.041 | 0.067 |
| Residual Std. Error | 16.836 (df = 7313) | 19.399 (df = 6789) | 17.928 (df = 6850) | 19.503 (df = 6655) | 17.623 (df = 7297) |

Estimates from OLS regressions of quality attribute ratings on status signal, examined above and below the median. Robust standard errors clustered at respondent level. + p<0.1; * p<0.05; ** p<0.01; *** p<0.001 for two-sided *t*-tests. Constant and controls not shown.

**Table S5.**

|  | Dependent variable: | | | | |
|---|---|---|---|---|---|
|  | overall.quality (1) | novelty (2) | validity (3) | generalizability (4) | significance (5) |
| percentile | 0.109*** (0.010) | 0.103*** (0.011) | 0.082*** (0.010) | 0.109*** (0.011) | 0.120*** (0.010) |
| status.signal | -5.035** (1.712) | -5.505** (2.091) | -3.504+ (1.887) | -6.722** (2.043) | -5.865** (1.842) |
| percentile X status.signal | 0.059* (0.023) | 0.071* (0.028) | 0.033 (0.026) | 0.070* (0.028) | 0.079** (0.025) |
| Controls | Y | Y | Y | Y | Y |
| Observations | 7,376 | 6,848 | 6,866 | 6,715 | 7,359 |
| $R^2$ | 0.087 | 0.047 | 0.045 | 0.048 | 0.079 |
| Adjusted $R^2$ | 0.083 | 0.043 | 0.041 | 0.044 | 0.075 |
| Residual Std. Error | 16.757 (df = 7343) | 19.382 (df = 6816) | 17.715 (df = 6834) | 19.476 (df = 6683) | 17.559 (df = 7327) |

Estimates from OLS regressions of quality attribute ratings on status signal and percentile. Robust standard errors clustered at respondent level. + $p<0.1$; * $p<0.05$; ** $p<0.01$; *** $p<0.001$ for two-sided *t*-tests. Constant and controls not shown.

**Table S6.**

|  | Dependent variable: | |
| --- | --- | --- |
|  | influence (1) | knowledge (2) |
| log.citations | 0.127*** (0.024) | 0.118*** (0.021) |
| expertise | 0.372*** (0.014) | 0.554*** (0.013) |
| first.paper | -0.276*** (0.019) | -0.230*** (0.016) |
| added.by.coauthor | -0.265*** (0.050) | -0.543*** (0.044) |
| Observations | 12,660 | 12,805 |
| $R^2$ | 0.743 | 0.811 |
| Adjusted $R^2$ | 0.388 | 0.553 |
| Residual Std. Error | 0.830 (df = 5307) | 0.736 (df = 5406) |

Estimates from OLS regressions of influence (model 1) and knowledge (model 2) on log-citations, author fixed effects, and controls. + p<0.1; * p<0.05; ** p<0.01; *** p<0.001 for two-sided *t*-tests.

**Table S7**

|                     | Dependent variable:       |
|---------------------|---------------------------|
|                     | major.influence           |
| log.citations       | 0.343*** |
|                     | (0.053)                   |
| expertise           | 0.802*** |
|                     | (0.038)                   |
| first.paper         | -0.691*** |
|                     | (0.061)                   |
| added.by.coauthor   | -0.705*** |
|                     | (0.130)                   |
| Observations        | 12,660                    |
| Log Likelihood      | -5,773.866                |
| Akaike Inf. Crit.   | 11,557.730                |
| Bayesian Inf. Crit. | 11,595.080                |

Estimates from generalized linear mixed model regression of *major.influence* on *log-citations*, author fixed effects, and controls. + p<0.1; * p<0.05; ** p<0.01; *** p<0.001 for two-sided *t*-tests.

**Table S8.**

|  | count | mean | std | min | 25% | 50% | 75% | max |
|---|---|---|---|---|---|---|---|---|
| percentile | 25476.0 | 67.69 | 25.39 | 2.0 | 49.00 | 73.0 | 90.00 | 99.00 |
| target_n_cites | 25476.0 | 81.83 | 795.68 | 1.0 | 9.00 | 20.0 | 48.00 | 77275.00 |
| target_log_cites | 25476.0 | 1.32 | 0.59 | 0.0 | 0.95 | 1.3 | 1.68 | 4.89 |
| influence | 17154.0 | 2.57 | 1.11 | 1.0 | 2.00 | 2.0 | 3.00 | 5.00 |
| top_influence | 17154.0 | 0.21 | 0.41 | 0.0 | 0.00 | 0.0 | 0.00 | 1.00 |
| knowledge | 17418.0 | 3.07 | 1.18 | 1.0 | 2.00 | 3.0 | 4.00 | 5.00 |
| overall_quality | 13177.0 | 69.41 | 17.32 | 1.0 | 60.00 | 71.0 | 81.00 | 99.00 |
| significance | 13084.0 | 70.55 | 17.97 | 1.0 | 60.00 | 72.0 | 83.00 | 99.00 |
| novelty | 12272.0 | 66.71 | 19.71 | 1.0 | 56.00 | 70.0 | 80.00 | 99.00 |
| validity | 12358.0 | 71.78 | 17.79 | 1.0 | 61.00 | 75.0 | 85.00 | 99.00 |
| generalizability | 11961.0 | 65.71 | 19.74 | 1.0 | 54.00 | 69.0 | 80.00 | 99.00 |
| gender | 18130.0 | 1.30 | 0.47 | 1.0 | 1.00 | 1.0 | 2.00 | 3.00 |
| position | 18212.0 | 4.26 | 2.72 | 1.0 | 2.00 | 3.0 | 7.00 | 9.00 |
| status_signal | 25476.0 | 0.15 | 0.36 | 0.0 | 0.00 | 0.0 | 0.00 | 1.00 |
| first_paper | 25476.0 | 0.50 | 0.50 | 0.0 | 0.00 | 0.5 | 1.00 | 1.00 |
| do_you_remember | 19528.0 | 1.10 | 0.30 | 1.0 | 1.00 | 1.0 | 1.00 | 2.00 |

Descriptive statistics of quantitative variables used in the dataset. Qualitative variables like *discipline* and *position* (professional rank) not included

**Table S9.**

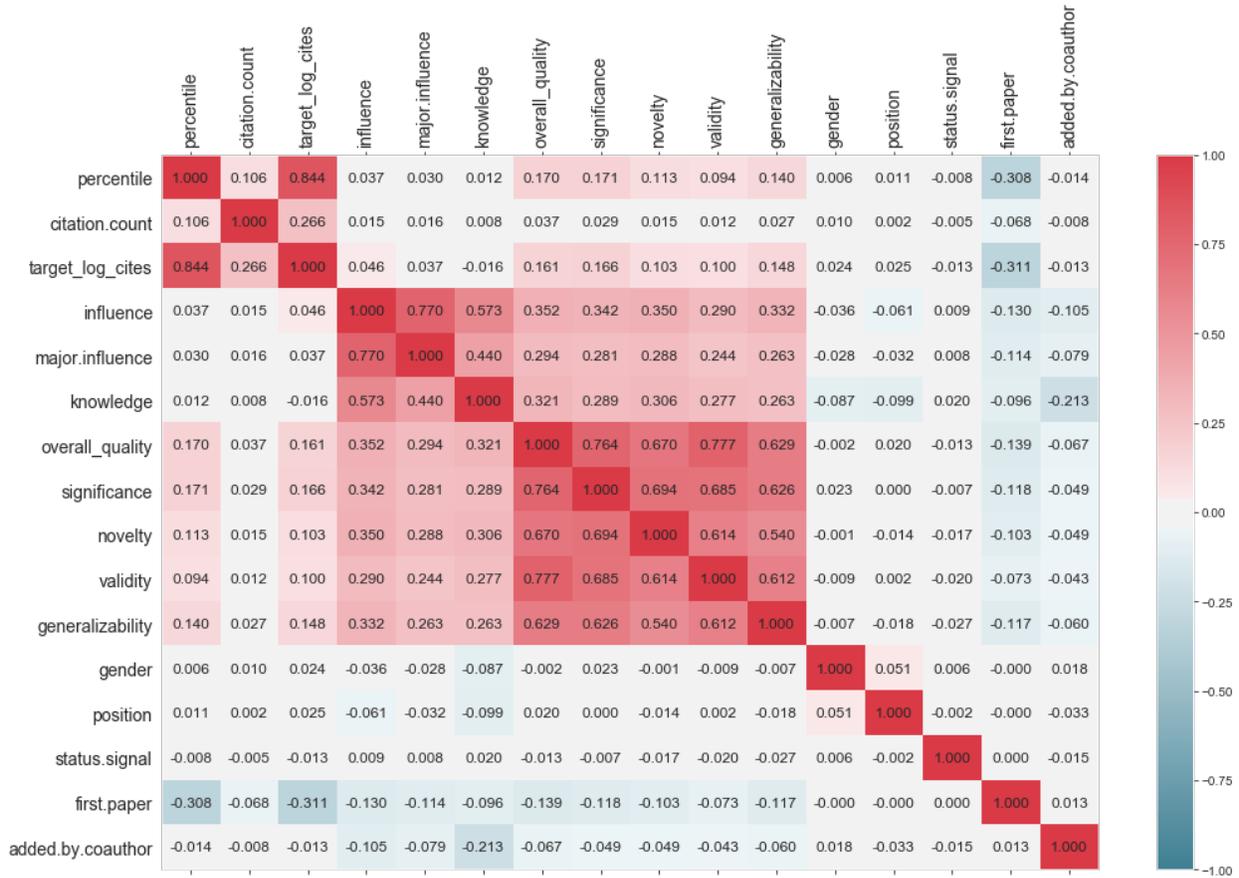

Correlation table. Qualitative variables like *discipline* are not included.